\documentclass[pra,twocolumn,amsmath,amssymb]{revtex4-1} %

\usepackage{epsfig,amsmath}
\usepackage{subfigure}
\usepackage{graphicx}% Include figure files
\usepackage{dcolumn}% Align table columns on decimal point
\usepackage{stmaryrd}
\usepackage{mathrsfs}
\usepackage{pifont}
\usepackage{amsthm}
\usepackage{amssymb}
\usepackage{bm}
\usepackage{latexsym}
\usepackage{hyperref}
\usepackage{color}

\newcommand{\la}{\langle}
\newcommand{\ra}{\rangle}

\newcommand{\ga}{\gamma}
\newcommand{\Ga}{\Gamma}

\newcommand{\al}{\alpha}
\newcommand{\si}{\sigma}

\newcommand{\non}{\nonumber}
\newcommand{\pa}{\partial}

\def\jpa#1{{ J.\ Phys.\ A} {\bf#1}}
\def\pra#1{{ Phys.\ Rev. A\/} {\bf#1}}
\def\prb#1{{ Phys.\ Rev. B\/} {\bf#1}}

\def\prl#1{{ Phys.\ Rev.\ Lett.} {\bf#1}}

\def\sci#1{{ Science} {\bf#1}}

\def\pla#1{{ Phys.\ Lett. A\/} {\bf#1}}
\def\rmp#1{{ Rev. \ Mod. \ Phys.} {\bf#1}}

\begin{document}

\title{Inverse Engineering Control in Open Quantum Systems}

\author{Jun Jing$^{1,2}$, Lian-Ao Wu$^{2}$ \footnote{Corresponding author: lianao.wu@ehu.es}, Marcelo S. Sarandy$^{3}$ and J. Gonzalo Muga$^{4,1}$ }

\affiliation{$^{1}$Department of Physics, Shanghai University, Shanghai 200444, China\\ $^{2}$Ikerbasque, Basque Foundation for Science, 48011 Bilbao and Department of Theoretical Physics and History of Science, The Basque Country University (UPV/EHU), PO Box 644, 48080 Bilbao, Spain \\ $^{3}$Instituto de F\'isica, Universidade Federal Fluminense, Av. Gal. Milton Tavares de Souza s/n, Gragoat\'a, 24210-346, Niter\'oi, RJ, Brazil\\ $^4$ Department of Physical Chemistry, The Basque Country University (UPV/EHU), PO Box 644, 48080 Bilbao, Spain }
%\{$^\ast$To whom correspondence should be addressed; Email:  jsmith@wherever.edu.}

\date{\today}

%%%%%%%%%%%%%%%%%%%%
\begin{abstract}
We propose a scheme for inverse engineering control in open quantum systems. Starting from an undetermined time evolution operator, a time-dependent Hamiltonian is derived in order to guide the system to attain an arbitrary target state at a predefined time. We calculate the fidelity of our inverse engineering control protocol in presence of the noise with respect to the stochastic fluctuation of the linear parameters of the Hamiltonian during the time evolution. For a special family of Hamiltonians for two-level systems, we show that the control evolution of the system under noise can be categorized into two standard decohering processes: dephasing and depolarization, for both Markovian and non-Markovian conditions. In particular, we illustrate our formalism by analysing the robustness of the engineered target state against errors. Moreover, we discuss the generalization of the inverse protocol for higher-dimensional systems.
\end{abstract}

\pacs{03.65.-w, 42.50.Lc, 42.50.Dv}

\maketitle

\section{Introduction}

To fulfill the requirement of high-precision quantum gates, teleportation, or state transfer, the aim of quantum state engineering (QSE)~\cite{Chuang,Adiab:Pass,Berry:09,Muga:11,Sarandy:11} is to manipulate the system and attain a target state, typically a pure state, at a designed time $T$; or more ambitiously, to drive the eigenstates of an initial Hamiltonian into those of a final Hamiltonian~\cite{Berry:09,Muga:11}. To be concrete, for a two-level system, a passage is constructed through which the system undergoes a transition from an initial state $|\psi (t=0)\rangle$ to a final state $|\psi(T)\rangle=\mu|1\rangle+\nu|0\rangle$, with $|\mu|^2+|\nu|^2=1$, in an undisturbed way. In a closed-system scenario, QSE has been considered through time optimal evolution~\cite{Carlini} and robust protocols of realizing QSE have been provided by adiabatic passages~\cite{Adiab:Pass,Bergmann-Kral} and by sequential convex programming~\cite{Robert}. However, for open systems~\cite{Breuer}, which are under environmental noise due to the onset of the system-bath dynamics~\cite{Gardiner}, QSE is significantly challenged in both theoretical and experimental implementations. Indeed, for decohering systems, there is a competition between the time required for adiabaticity and the decoherence time scales~\cite{Sarandy:05}, which may limit the applicability of the adiabatic
approach~\cite{Martin:12}. Identifying protocols that are both fast and fault-tolerant is therefore an important research direction for quantum information processing and quantum control~\cite{Muga:11}.

A non-adiabatic approach for QSE~\cite{Muga:11,Sarandy:11,Wan}, originally defined in the context of closed systems, is the inverse engineering control based on Lewis-Riesenfeld invariants~\cite{LR}. In this approach, the quantum dynamics is dictated by a dynamical invariant $I(t)$, which is an operator defined through the von Neumann equation, $\frac{\pa}{\pa t}I(t) + i \left[ H(t), I(t) \right]=0$ (setting $\hbar=1$). By denoting the instantaneous eigenbasis of $I(t)$ as an orthonormal set $\{|\phi_n(t)\rangle\}$, the method of quantum control by invariants first considers the initial state $|\psi (t=0)\rangle$ to be one of the eigenstates of $I(t)$, say $|\phi_0(0)\rangle$, and the target state to be $|\phi_0(T)\rangle$. The operator $I(t)$ is then designed to conveniently interpolate between $|\phi_0(0)\rangle$ and $|\phi_0(T)\rangle$. In particular, the von Neumann equation implies that $I(t)=U(t)I(0)U^\dagger(t)$, where $U(t)$ denotes the time evolution operator. Then, $U(t)$ will satisfy $U(T) |\phi_0(0)\rangle = |\phi_0(T)\rangle$, i.e. a system that is initially prepared in a given eigenlevel of $I(t)$ will be kept in the corresponding instantaneous eigenlevel for any time $T$, yielding a non-transitional evolution. Hence, QSE by inverse control can be implemented by obtaining a desired evolution operator $U(t)$  from $I(t)$, which will give rise in a further step to a time-dependent control Hamiltonian able to physically generate the engineered target state $|\phi_0(T)\rangle$ at time $T$.

The aim of this work is to propose a scheme to investigate QSE by inverse control in presence of decoherence. This strategy is different from the usual techniques of protecting the system against the external environmental noise through fast pulse sequences, such as bang-bang control~\cite{Lloyd} and optimal control~\cite{Schirmer}. Moreover, it is also distinct from the direct kinematic controllability of open quantum systems~\cite{Herschel}. The starting point of our framework is the observation that the von Neumann equation implies that the density operator $\rho(t)$ is a Lewis-Riesenfeld invariant by itself. Indeed, $\rho(t)$ satisfies the same equation of motion as $I(t)$, namely, $\frac{\pa}{\pa t}\rho+i[H, \rho]=0$. Then, $I(t)$ and $\rho(t)$ share the same eigenbasis $\{|\phi_n(t)\rangle\}$, and $\rho(t)=U(t)\rho(0)U^\dagger(t)$ where $\rho(0)$ is an arbitrary initial density matrix. For the evolution operator $U(t)$, we obtain $U(t)=\sum_n |\phi_n(t)\rangle\langle\phi_n(0)|$. Therefore, $\rho(t)$ and $U(t)$ are mutually equivalent to each other without considering the Lewis-Riesenfeld phase~\cite{LR}. Hence, the procedure is firstly to get the dynamics of the system as expressed by the time evolution operator $U(t)$. As a second step, we inversely obtain the Hamiltonian $H(t)=i\dot{U}U^\dag$. Noise is then introduced through stochastic fluctuations in the Hamiltonian of the system. Such fluctuations will be identified according to specific quantum channels~\cite{Chuang}, which are physically defined as some special pipelines intended to carry quantum information through an open quantum system. In particular, we find the conditions in which the rapid and robust state passage or population transfer could be realized and those in which the QSE ends up with a mixed state.

\section{Inverse engineering theory in quantum system control}

We start from a general time evolution operator for a two-level system,
\begin{equation}
U(t)=\cos\theta(t)+i\sin\theta(t)\vec{\si}\cdot\vec{n}(t),
\end{equation}
where $\vec{\si} = (\sigma_x,\sigma_y,\sigma_z)$ represents the vector of Pauli operator and $\vec{n}$ denotes a unit vector in space. The Hamiltonian could be expressed as \cite{Wu93}
\begin{eqnarray}\label{H}
H=i\dot{U}U^\dag=-\vec{\si}\cdot[\dot{\theta}\vec{n}
+\sin\theta\cos\theta\dot{\vec{n}}
+\sin^2\theta(\dot{\vec{n}}\times\vec{n})].
\end{eqnarray}
Consider a general unit vector $\vec{n}=\cos\alpha\vec{x}+\sin\alpha\cos\beta\vec{y}+\sin\alpha\sin\beta\vec{z}$, and substitute it into Eq.~(\ref{H}), then the Hamiltonian reads
\begin{eqnarray}\non
H&=&\si_x(-\dot{\theta}\cos\alpha+\dot{\alpha}\sin\theta\cos\theta\sin\alpha
+\dot{\beta}\sin^2\theta\sin^2\alpha) \\ \non &+& \si_y[-\dot{\theta}\sin\alpha\cos\beta \\ \non
&-&\dot{\alpha}\sin\theta(\cos\theta\cos\alpha\cos\beta
+\sin\theta\sin\beta) \\ \non
&+&\dot{\beta}\sin\theta\sin\alpha(\cos\theta\sin\beta
-\sin\theta\cos\alpha\cos\beta)] \\ \non &+& \si_z[-\dot{\theta}\sin\alpha\sin\beta \\ \non
&-&\dot{\alpha}\sin\theta(\cos\theta\cos\alpha\sin\beta-\sin\theta\cos\beta)
\\ \label{H1} &-&\dot{\beta}\sin\theta\sin\alpha(\cos\theta\cos\beta
+\sin\theta\cos\alpha\sin\beta)].
\end{eqnarray}
The Hamiltonian $H$ in Eq.~(\ref{H1}) is an explicit function of six variables, namely, $H=H(\dot{\alpha},\dot{\theta},\dot{\beta},\alpha,\theta,\beta)$. In particular, $H$ is linear in the variables $\dot{\alpha}$, $\dot{\theta}$, and $\dot{\beta}$, and nonlinear in the variables $\alpha$, $\theta$ and $\beta$. Within an open-system dynamics, any of these variables could be associated with a source of stochastic noise, e.g. the semi-classical dephasing model~\cite{YE}. We will be interested here in the fluctuations of the linear variables.

For the sake of simplicity, we will take $\beta=0$ to illustrate our theory. In this case, $H$ and $U$ read
\begin{eqnarray} \non
H&=&\si_x(-\dot{\theta}\cos\alpha+\dot{\alpha}\sin\theta\cos\theta\sin\alpha)
\\  \label{H2} &-&\si_y(\dot{\theta}\sin\alpha+\dot{\alpha}\sin\theta\cos\theta\cos\alpha)
+\si_z\dot{\alpha}\sin^2\theta, \\ U&=&\cos\theta+i\sin\theta(\si_x\cos\alpha+\si_y\sin\alpha).
\end{eqnarray}
In the absence of noise, when the system is prepared as $|\psi(0)\ra=|0\ra$, it will evolve into
\begin{equation}\label{Ut}
U|0\ra\la0|U^\dag=\left(\begin{array}{cc}
       \sin^2\theta & i\sin\theta\cos\theta e^{-i\alpha} \\
       -i\sin\theta\cos\theta e^{i\alpha} & \cos^2\theta
    \end{array}\right),
\end{equation}
which is equivalent to a pure state  $|\psi(t)\ra=\sin\theta(t)|1\ra-ie^{i\alpha(t)}\cos\theta(t)|0\ra$. Here $|0\ra$ and $|1\ra$ represent the two eigenvectors of $\si_z$, with eigenvalues $-1$ and $1$, respectively. Besides the consistency condition $\sin\theta(0)=0$, there is almost no limit on the choices of functions $\theta(t)$ and $\alpha(t)$ for engineering the system to reach an arbitrary target state $\mu|1\ra+\nu|0\ra$ at arbitrary time $T$, as long as $\sin\theta(T)=\mu$ and $-ie^{i\alpha(T)}\cos\theta(T)=\nu$. For example, if $\mu=\nu=\frac{1}{\sqrt{2}}$, then a simple choice is $\theta(t)=\frac{\pi t}{4T}$, $\alpha=0$, and $H=-\frac{\pi}{4T}\si_x$. This control dynamics allows for a target state achieved at an optimal time $T$ compatible with the brachistochrone solution~\cite{Carlini}.

By adding decoherence to the system, the passage described above is not always attainable. In order to develop a protocol for QSE under the effect of decoherence, we will individually consider the special family of Hamiltonians given by Eq.~(\ref{H2}) under the effect of stochastic fluctuations over the linear variables $\dot{\alpha}$ and $\dot{\theta}$. A more general setup could be considered by investigating the cases of correlated noise or the noise attached to the nonlinear variables of the Hamiltonian. In any case, the approach introduced in this work can also be generalized to these situations. In presence of noise, the density matrix $\rho(t)$ at time $t$ can be evaluated by the ensemble average over different realizations of time evolution operator with fluctuating parameters due to noise, i.e.
\begin{equation}
\rho(t)\equiv M[U(\xi)|\psi(0)\ra\la\psi(0)|U^\dag(\xi)],
\label{ensemble}
\end{equation}
where $M[\cdot]$ denotes ensemble average, $\xi$ the noise parameter, and $U(\xi)$ the instantaneous time evolution operator (\ref{Ut}). To evaluate the quality of QSE under the effect of noise, we adopt the control fidelity, which is defined by
\begin{equation}
 \mathcal{F}(t)\equiv\sqrt{\la\psi_0(t)|\rho(t)|\psi_0(t)\ra},
\label{fidelity}
\end{equation}
where $|\psi_0(t)\rangle$ is the engineered state without noise. The non-unitary dynamics implied by Eq.~(\ref{ensemble}) can be identified with different quantum channels. By identifying prototypical quantum channels, we can illustrate the state preparation by the inverse engineering procedure.

\subsection{State preparation under dephasing}

Suppose $\dot{\alpha}$ is not a stable variable, i.e. $\dot{\alpha}=a(t)+\xi(t)$, where $\xi(t)$ is chosen as Gaussian noise. Then, ensemble average yields $M[\dot{\alpha}]=a(t)+\xi_0$ and $M[(\xi(t)-\xi_0)(\xi(s)-\xi_0)]=g(t,s)$. Here $\xi_0$ is the mean value (or the bias) of $\xi(t)$. When $\xi_0=0$, the noise is categorized into a non-biased noise. The quantity $g(t,s)$ is the correlation function for the non-biased Gaussian noise. In order to compute the mean value over Eq.~(\ref{Ut}), all of the terms involving $\theta$ are kept unchanged, but
$M[e^{\pm i\alpha(t)}]=e^{-\frac{1}{2}\int_0^tG(s)ds}e^{\pm i[\alpha_0(t)+\xi_0t]}$, where $\alpha_0(t)\equiv\int_0^ta(s)ds$ and $G(s)\equiv\int_0^sg(s,s')ds'$. Therefore the density matrix is
\begin{equation}\label{ma}
\rho(t)=\left(\begin{array}{cc}
       \frac{1-\cos(2\theta)}{2} & \frac{i\sin(2\theta)re^{-i[\alpha_0(t)+\xi_0t]}}{2} \\
       \frac{-i\sin(2\theta)re^{i[\alpha_0(t)+\xi_0t]}}{2} &  \frac{1+\cos(2\theta)}{2}
    \end{array}\right),
\end{equation}
where $r\equiv r(t)=e^{-\frac{1}{2}\int_0^tG(s)ds}$ is a time-dependent noise factor determined by the noise correlation function. For any stationary noise, $r(0)=1\geq r(t)\geq r(\infty)=0$. We observe that this stochastic fluctuation over $\dot{\alpha}$ induces a purely dephasing process, i.e. the diagonal terms are not influenced by noise and the off-diagonal terms are damped at the same rate in the density matrix, which will end up with a mixed state $\sin^2\theta|1\ra\la1|+\cos^2\theta|0\ra\la0|$ in the asymptotic time limit. If we use $\rho_0\equiv|\psi_0\ra\la\psi_0|$ to represent the initial density matrix free of noise, as given by Eq.~(\ref{Ut}), then Eq.~(\ref{ma}) can be expressed in Kraus operator-sum representation \cite{Chuang,Kraus,YuEberly} for a qubit under dephasing, $\rho(t)=K_1\rho_0K_1^\dag+K_2\rho_0K_2^\dag$, where $K_1={\rm diag}[re^{-i\xi_0t}, 1]$ and $K_2={\rm diag}[\sqrt{1-|r|^2}, 0]$. ${\rm diag}[\cdot]$ represents the diagonal matrix. Concerning the control fidelity, the evaluation of Eq.~(\ref{fidelity}) yields
\begin{equation}\label{Falpha}
\mathcal{F}(t,\dot{\alpha})=\sqrt{1-0.5\sin^2(2\theta)(1-r\cos\xi_0t)}.
\end{equation}
In this model, the control fidelity is sensitive to the target state. In particular, $\mathcal{F}\geq\sqrt{1-\frac{\sin^2(2\theta)}{2}}$, i.e. it is increased by decreasing $\theta$, $0\leq\theta\leq\pi/4$. Also the noise bias $\xi_0\neq0$ deteriorates the fidelity, unless the time $T$ chosen to arrive the target state satisfies $\cos(\xi_0T)=0$. Remarkably, the result with noise is meaningful for pure state control. When the target state is $|1\ra$, it is still achievable at time $T$ as long as $\theta(T)=\frac{(2k+1)\pi}{2}$, $k=0,1,2,\cdots$. Thus a complete population inversion~\cite{Herschel,Wang} could be realized, which is robust to the dephasing noise from $\dot{\alpha}$.

On the other hand, for $\xi_0=0$ and $r(t)$ representing a slowly varying function of time, we can still attain a target state with high fidelity in a short time $T$. For instance, suppose $r(t)=\exp(-\Ga t/4)$, which corresponds to Markovian white noise, with $g(t,s)=\Ga\delta(t,s)$ and $\Ga$ denoting the coupling strength with the environment. At the same time, suppose that it is required that $\mathcal{F}\geq\mathcal{F}_c$, a threshold value for quantum information processing. Then, $T$ must be less than a critical time $t_c$, which is given by
\begin{equation}
t_c=-\frac{4}{\Ga}\ln\left[1-\frac{2(1-\mathcal{F}_c^2)}{\sin^2(2\theta)}\right].
\end{equation}
If $\sin^2(2\theta)=0.5$ and $\mathcal{F}_c=0.99$, then $\Ga t_c\approx0.33$. For Rydberg atoms in a cavity or electron spins in quantum dots \cite{You}, we have $\Ga\sim10^6~{\rm Hz}$ and $T\sim0.33~\mu s$. It is indeed a very hard challenge to the external control time-dependent fields, such as laser pulse sequences.

\begin{figure}[htbp]
\centering
\subfigure{\label{Fa1}
\includegraphics[width=3in]{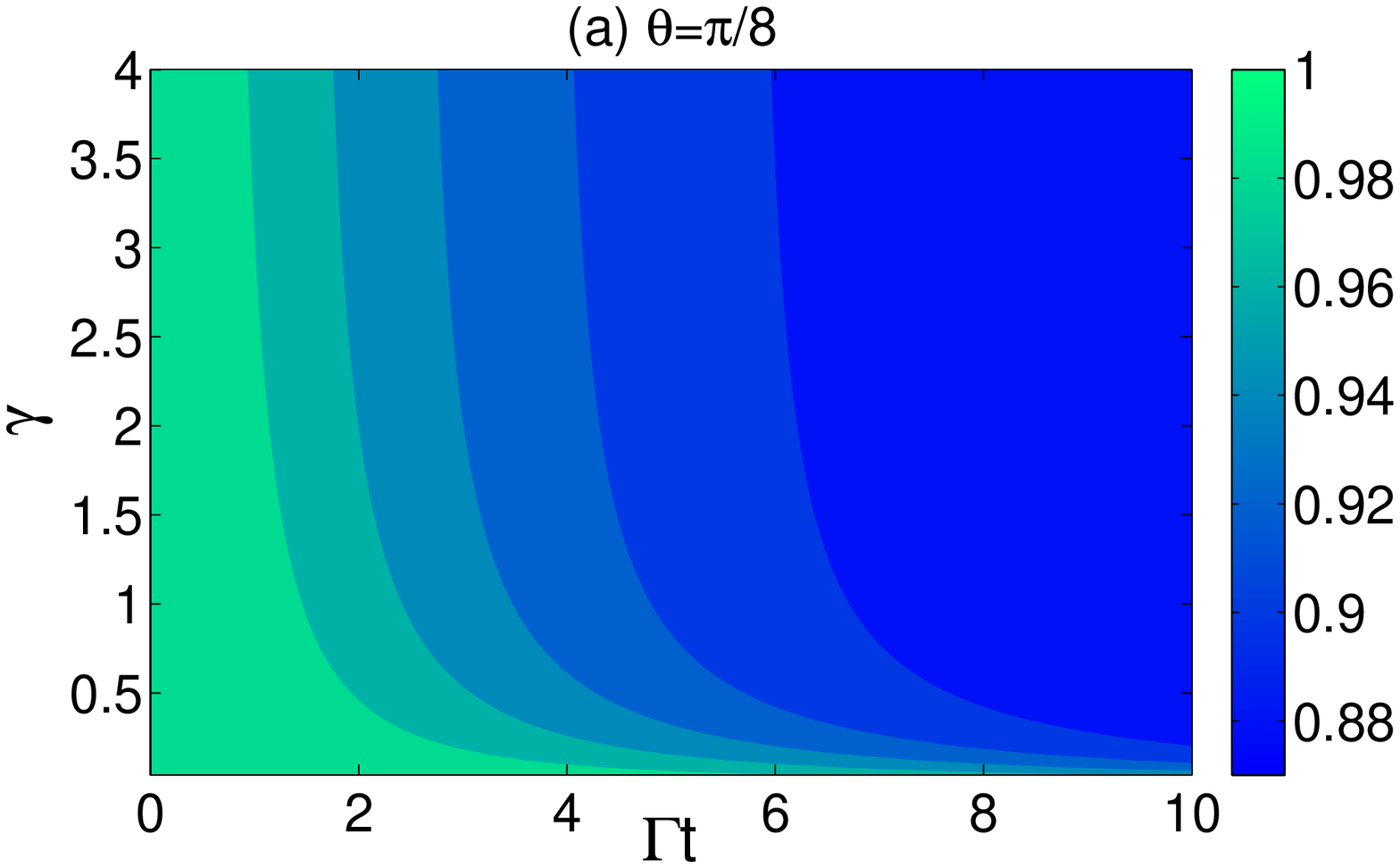}}
\subfigure{\label{Fa2}
\includegraphics[width=3in]{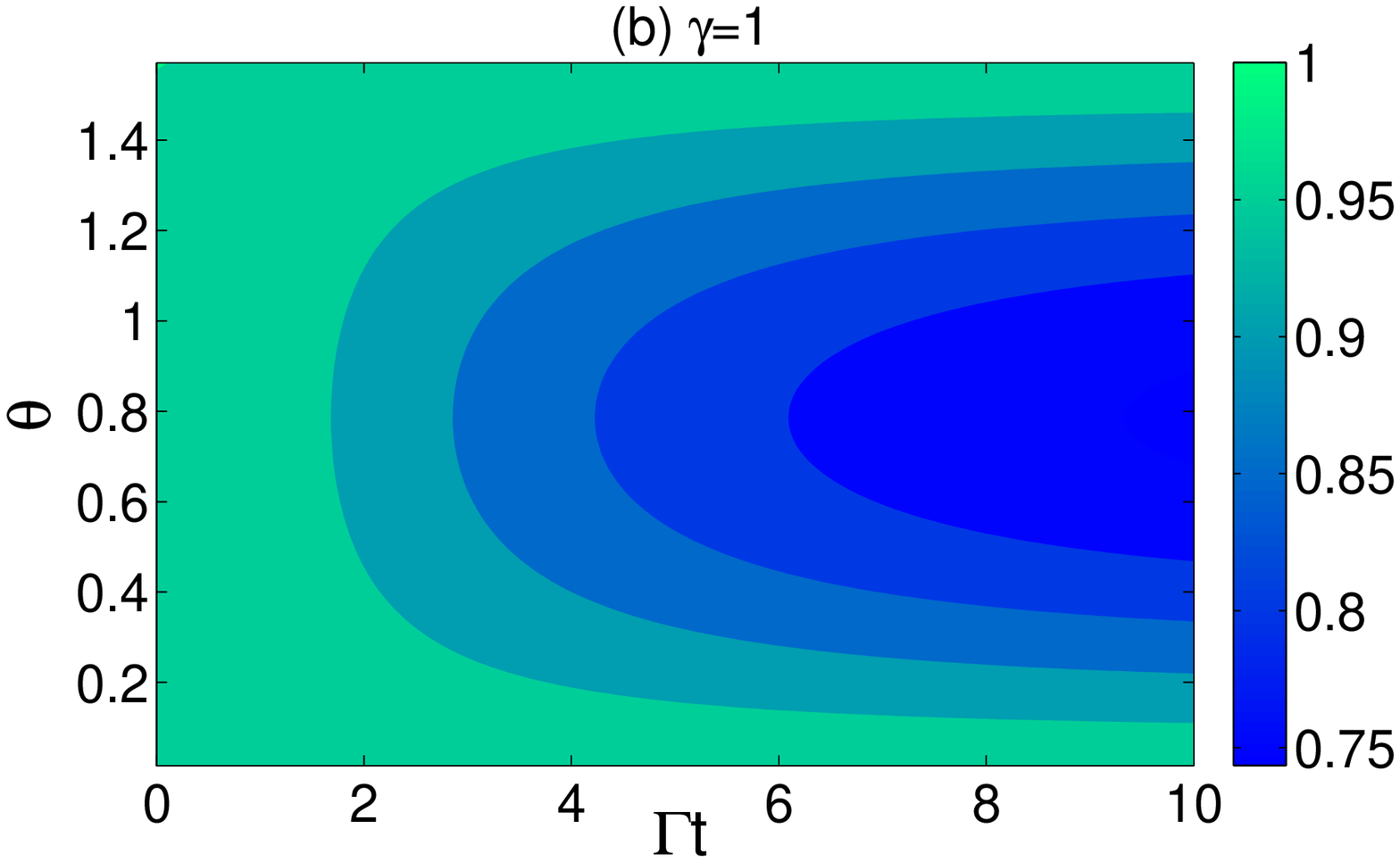}}
\caption{(Color online) Fidelity $\mathcal{F}$ under non-Markovian dephasing, with $\xi_0=0$, as a function of $\Gamma t$ (dimensionless), $\gamma$ (in the unit of frequency), and $\theta$ (angle in radian measure). (a) The target state is chosen as $\theta=\pi/8$; (b) The noise parameter is chosen as $\ga=1$. }\label{Fa}
\end{figure}

However, if the noise is not a Markovian white noise, then the time for QSE would be greatly extended. Here we use the Ornstein-Uhlenbeck process with the correlation function $g(t,s)=\Ga\ga e^{-\ga|t-s|}/2$, to demonstrate the effect of non-Markovian noise, where the time-dependent damping function is $r(t)=\exp\{-\Ga[t+(e^{-\ga t}-1)/\ga]/4\}$. When $\ga\rightarrow\infty$, $r(t)$ reduces to the exponential decay function as that in the case of Markovian noise. On the other hand, for $\ga\rightarrow0$, the decay rate of $r$ is intensively suppressed, which means the fidelity maintains a value very close to unity in a much longer time interval. In Fig. \ref{Fa}, we plot the dynamics of the control fidelity under the Ornstein-Uhlenbeck noise. Figure \ref{Fa1} shows the effect of $\ga$ with a fixed value of $\theta=\pi/8$, i.e. $\sin^2(2\theta)=0.5$. In a strong non-Markovian regime $\ga=0.1$, the fidelity could be kept not less than $0.99$ until $\Ga t\approx8$, i.e. $T\leq8~\mu s$ for atoms in a cavity, which is almost $24$ times as what we get with the Markov noise. Figure \ref{Fa2} shows the fidelity dependence on different target state with $0\leq\theta\leq\pi/2$. For a target state $|\psi\ra=\sin\theta|1\ra-ie^{i\alpha}\cos\theta|0\ra$, the population at the high-level $|1\ra$ is $\sin^2\theta$ that is monotonic in this scale. Yet the plot is symmetric with respect to $\theta=\pi/4$, where the survival time for high fidelity QSE takes a minimal value. This means that the difficulty in the state control is not proportional to the state population.

\subsection{State preparation under depolarization}

In this model, it is assumed that $\dot{\theta}=b(t)+\xi(t)$. By employing the same notation as in the previous analysis, we obtain, after ensemble average, that the density matrix is given by
\begin{equation}\label{dep}
\rho(t)=\left(\begin{array}{cc}
       \frac{1-r^4\cos[2\theta_0(t)+2\xi_0t]}{2} & \frac{ir^4\sin[2\theta_0(t)+2\xi_0t]e^{-i\alpha}}{2} \\
       \frac{-ir^4\sin[2\theta_0(t)+2\xi_0t]e^{i\alpha}}{2} & \frac{1+r^4\cos[2\theta_0(t)+2\xi_0t]}{2}
    \end{array}\right),
\end{equation}
where $\theta_0(t)\equiv\int_0^tb(s)ds$. The final state of $\rho(t)$ is the most mixed state $\frac{1}{2}\mathcal{I}$, where $\mathcal{I}$ is the identity matrix. Subsequently, the fidelity is
\begin{equation}\label{Ftheta}
\mathcal{F}(t,\dot{\theta})=\sqrt{0.5[1+r^4\cos(2\xi_0t)]},
\end{equation}
which means that, in this model, the fidelity is independent of the choice of  target state. This is a dramatic difference between the noises induced by $\dot{\theta}$ and $\dot{\alpha}$.

\begin{figure}[htbp]
\centering
\includegraphics[width=3in]{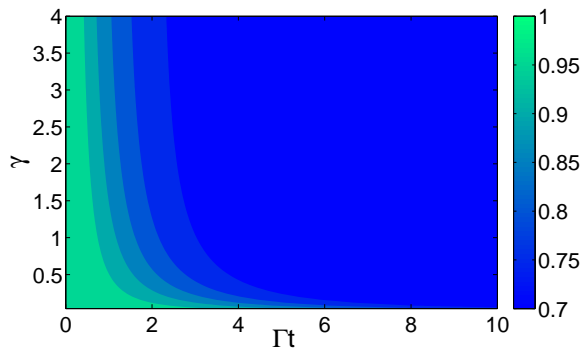}
\caption{(Color online) The fidelity $\mathcal{F}$ dynamics under a non-Markovian depolarization noise ($\xi_0=0$) with different $\ga$ (in the unit of frequency). }\label{Ft}
\end{figure}

When $\xi_0=0$, the external noise in $\dot{\theta}$ induces a standard depolarization quantum channel~\cite{Chuang} for the open system, since under influence of the noise, $\rho(t)=\sum_{j=1}^4D_j\rho_0D_j^\dag$, where the Kraus operators are $D_1=\sqrt{1+3r^4}\mathcal{I}/2$, $D_2=\sqrt{1-r^4}\si_x/2$, $D_3=\sqrt{1-r^4}\si_y/2$, $D_4=\sqrt{1-r^4}\si_z/2$. Here $r$ measures the probability that the system is undisturbed by the noise. Note that this process constitutes a novel quantum microscopic model for depolarization dynamics.

When $\xi_0\neq0$, there is a subtle aspect left for the initial state preparation. Indeed, in Eq.~(\ref{Ut}), $|\psi(0)\ra=|0\ra$ means $\sin\theta_0(0)=0$ for closed systems. However, for open systems, the initial state is chosen as $\sin\theta(0)=\sin[\theta_0(t)+\xi_0]=\sin\xi_0\neq0$ in presence of noise. This can be resolved by measuring the noise bias through a spectroscopy analysis in an experimental implementation.

By taking a Markovian noise with $r(t)=\exp(-\Ga t/4)$ and $\xi_0=0$, the critical time $t_c$ for control is
\begin{equation}
t_c=-\ln(2\mathcal{F}_c^2-1)/\Ga.
\end{equation}
If $\mathcal{F}_c=0.99$, then $\Ga t=0.04$, which is valid for an arbitrary target state.

The control fidelity under the non-Markovian depolarization noise is demonstrated in Fig. \ref{Ft}. When $\ga=0.1$, the fidelity could be maintained as $0.99$ until $\Ga t\approx3$, i.e. $T\leq3\mu s$ for atoms in a cavity, which is about $75$ times as that in the case of white noise. A comparison of Eqs.~(\ref{Falpha}) and (\ref{Ftheta}) shows that the depolarization noise plays a more severe role on destroying QSE. In principle, there is no chance to perfectly attain an arbitrary pure target state or realize a perfect population transfer with depolarization noise even in the non-Markovian regime.

\section{Discussions}

A first important observation is that, for either dephasing induced by $\dot{\alpha}$ or depolarization induced by $\dot{\theta}$, the category description of quantum channel for two-level systems holds irrespectively to the statistical properties of the noise. Distinct noises defined by different correlation functions will lead to different damping functions $r(t)$, which will determine the control fidelity. Concerning the critical fidelity, for a quantum system embedded into a non-Markovian environment, $r(t)$ could be preserved to nearly unity for a long time. Therefore, non-Markovianity is a useful tool to achieve a suitable control in QSE.

Moreover, this universal inverse QSE inverse protocol for two-level systems can be extended straightforwardly into higher-dimensional systems. Consider the stimulated Raman adiabatic passage (STIRAP)~\cite{Bergmann:98} in a three-level atomic system, which targets on the population transition between
$|0\ra$ and $|2\ra$ without disturbing the quasi-stable state $|1\ra$. The eigenbasis could be chosen as $|\phi_1\ra=\sin\theta\sin\al|0\ra+\cos\al|1\ra+\cos\theta\sin\al|2\ra$, $|\phi_2\ra=\sin\theta\cos\al|0\ra-\sin\al|1\ra+\cos\theta\cos\al|2\ra$, and $|\phi_3\ra=\cos\theta|0\ra-\sin\theta|2\ra$. Assume $\theta(0)=\al(0)=0$.  Then $U$ and $H$ are obtained as
\begin{eqnarray}
U(t)&=&\left(\begin{array}{ccc}
\cos\theta\cos\al & \cos\theta\sin\al & -\sin\theta \\
 -\sin\al & \cos\al & 0 \\
\sin\theta\cos\al & \sin\theta\sin\al & \cos\theta
\end{array}\right), \\
H(t)&=&i\left(\begin{array}{ccc}
0 & \dot{\al}\cos\theta & -\dot{\theta} \\
-\dot{\al}\cos\theta & 0 & -\dot{\al}\sin\theta \\
\dot{\theta} & \dot{\al}\sin\theta & 0
\end{array}\right).
\end{eqnarray}
As expected, it demonstrates that the linear parameters of Hamiltonian are $\dot{\al}$ and $\dot{\theta}$. If the system is prepared as $|\psi(0)\ra=|0\ra$, then $|\psi(t)\ra=-\sin\theta(t)|2\ra+\cos\theta(t)|0\ra$ in case of a closed system. That would be an ideal STIRAP. Yet, with linear noise $\dot{\theta}=b(t)+\xi(t)$ (here we choose $\xi_0=0$), the density matrix becomes
\begin{equation}
\rho(t)=\left(\begin{array}{ccc}
       \frac{1-r^4\cos[2\theta_0(t)]}{2} & 0 & \frac{-r^4\sin[2\theta_0(t)]}{2} \\
       0 & 0 & 0 \\
       \frac{-r^4\sin[2\theta_0(t)]}{2} & 0 & \frac{1+r^4\cos[2\theta_0(t)]}{2}
    \end{array}\right),
\end{equation}
which is almost the same as Eq.~(\ref{dep}) with $\al=\xi_0=0$, i.e. it is also a depolarization process. Then the control fidelity is given by $\sqrt{(1+r^4)/2}$, which means the noise will definitely destroy the perfect QSE. However, beyond the STIRAP, if the state is prepared as $|1\ra$ and the target state is $\mu|2\ra+\nu|0\ra$, the fidelity under the {\it same} noise is $\mathcal{F}=\sqrt{\frac{1+r^4}{2}\sin^4\al+2r\sin^2\al\cos^2\al+\cos^4\al}$, which is independent of both the target state and $\theta_0$. When $\al(T)=0$, $\mathcal{F}$ could approach unity, meaning that now the QSE is robust against the $\dot{\theta}$ noise.

In comparison with other recent Markovian approaches to deal with decoherence due to weak system-bath interactions~\cite{Muga:11}, our inverse QSE method has no limitation on the correlation function of the errors in the system parameters, which can imply in more general sources of environmental noises, such as non-Markovian evolutions and strong system-bath interactions. In this scenario, pure dephasing and depolarization appear as two important quantum channels associated with the fluctuations in the linear parameters of the engineered Hamiltonian, but other sources of errors could be considered. Moreover, following the strategy presented in this section, systems with more than two or three levels could be investigated.

\section{Conclusion}

In conclusion, we have investigated decoherence in the linear parameters of the Hamiltonians of two-level and three-level open quantum systems, discussing its consequences for inverse QSE protocol. Exact results have been expressed by different standard quantum channels, each of them imposing different restrictions on the control fidelity. Non-Markovian noise is found to be helpful to attain a high fidelity state passage for a much longer time than the Markov white noise. Generalization of our approach for other higher-dimensional systems is a promising route to reveal more abundant relations between control and noise.

\acknowledgments
We acknowledge grant support from the NSFC No. 11175110, the Basque Government (grant IT472-10), the Spanish MICINN (Projects No. FIS2012-36673-C03-01 and -03), the UPV/EHU program UFI 11/55-01-2013, the Brazilian agencies CNPq, FAPERJ, and the Brazilian National Institute for Science and Technology of Quantum Information (INCT-IQ).

\end{document}